\documentclass{jpsj-suppl}
\usepackage{txfonts}
\usepackage[dvipdfmx]{graphicx}
\usepackage{amsmath}
\usepackage{multirow}
\usepackage{braket}

\newcommand{\nuc}{n^{\mathrm{U}}_c}
\newcommand{\nlc}{n^{\mathrm{L}}_c}
\newcommand{\npc}{n^+_c}
\newcommand{\nmc}{n^-_c}

\begin{document}

\title{
Path-integral Monte Carlo for the Gauge-fixed Berry Connection and the Local $Z_2$ Berry Phase
}

\author{Yuichi \textsc{Motoyama}$^1$ and Synge \textsc{Todo}$^2$}
\inst{$^1$Department of Applied Physics, University of Tokyo, Tokyo 113-8656, Japan \\
$^2$Institute for Solid State Physics, University of Tokyo, 7-1-26-R501 Port Island South, Kobe 650-0047, Japan}
\email{yomichi@looper.t.u-tokyo.ac.jp}

\recdate{July 15, 2013}

\abst{
We present a general formula of the gauge-fixed Berry connection
which can be evaluated by path-integral Monte Carlo method.
We also propose that the gauge-fixed local Berry connection can be another
effective tool to estimate precisely the quantum critical point.
For a demonstration, we calculate the gauge-fixed Berry connection and the local $Z_2$ Berry phase
of the antiferromagnetic Heisenberg model
on a staggered bond-alternating ladder,
and estimated quantum critical point is consistent with other methods.
}

\kword{quantum spin system, quantum phase transition, Berry phase, quantum Monte Carlo, sign problem}

\maketitle
\section{Introduction}
The local $Z_2$ Berry phase, proposed by Hatsugai~\cite{Hatsugai2004},
is a topological order parameter
that can characterize a short-range entanglement state such as a spin singlet pair.
This is just a Berry phase that the ground state obtains
under the local perturbation, for example, only on one bond term.
If a model has some symmetry in the space of degree of freedom,
this symmetry quantizes the Berry phase.
A value of the local Berry phase depends on where the local perturbation works on.
For the $S=1/2$ antiferromagnetic Heisenberg model, for example, the $Z_2$ spin inversion symmetry (time reversal symmetry) quantizes
the Berry phase to $0$ or $\pi$ mod $2\pi$.
When the local perturbation is bond twist, replacing a bond Hamiltonian on $\braket{kl}$ bond
$S_k^+S_l^- + S_k^-S_l^+$ by $e^{i\theta}S_k^+S_l^- + e^{-i\theta}S_k^-S_l^+,$
the local Berry phase will be $\pi$ or $0$ dependending on whether the twisted bond is a valence bond or not
in the valence bond solid picture, respectively.
Since this quantization is protected by the symmetry,
any other perturbation cannot change the Berry phase as long as the energy gap above the ground state remains finite.
Therefore, changes of the spatial pattern of $\pi$-valued local quantized Berry phase enable us to catch quantum critical points.

Since the local Berry phase has been calculated only by the exact diagonalization method,
the reachable system size is strongly limited especially in higher dimensions
and finite size effects remain.
Thus, we developed a non-biased large-scale quantum Monte Carlo method for the local quantized Berry phase~\cite{MotoyamaT2013}.
We applied the method to an antiferromagnetic Heisenberg model on a staggered bond-alternating ladder~\cite{MotoyamaT2013}.

\section{Monte Carlo for the Berry connection}
In the following, we will derive the evaluation form of the Berry connection
by path-integral Monte Carlo method
in more general way than one that the authors derived before in the past rapid communication~\cite{MotoyamaT2013}.
The ground state of a Hamiltonian $\mathcal{H}(\theta)$ is given by the projection method as
$\ket{\Psi(\theta,\sigma)} = \lim_{\beta\to\infty}\ket{\Psi(\theta, \sigma, \beta)} = \lim_{\beta\to\infty}N(\theta,\sigma, \beta)e^{-\beta\mathcal{H}(\theta)/2}\ket{\sigma}$,
where $\ket{\sigma}$ is an arbitrary initial state which is not orthogonal to the ground state,
$\beta > 0$ is a projection parameter,
and $N \in \mathbf{R}$ is a normalization factor, $\braket{\Psi(\theta,\sigma,\beta)|\Psi(\theta,\sigma,\beta)} = 1.$
Note that $\ket{\sigma}$ (more strictly speaking, $\ket{\sigma(\theta)}$) fixes the phase of the obtained ground state.
By using the path-integral expansion,
the inner product between the ground states of two Hamiltonians, $\mathcal{H}(\theta)$ and $\mathcal{H}(\phi)$, and the normalization factor can be written in terms of world-line representation as
\begin{equation}
\begin{split}
\tilde{A}(\theta, \phi, \sigma, \beta) &\equiv  \Braket{\Psi(\theta, \sigma, \beta)|\Psi(\phi, \sigma, \beta)} \\
&= N(\theta, \sigma, \beta)N(\phi,\sigma,\beta) \Braket{\sigma | e^{-\beta\mathcal{H}(\theta)/2}e^{-\beta\mathcal{H}(\phi)/2}|\sigma} \\ 
&= N(\theta, \sigma, \beta)N(\phi,\sigma,\beta) \sum_c W(c, \theta, \phi, \sigma, \beta) \\
&\equiv N_\theta N_\phi \sum_c W(c,\theta,\phi)
\end{split}
\label{eq:inner-product}
\end{equation}
with $N_\theta^{-2} = \sum_c W(c,\theta,\theta)$, where $c$ is the index of world-line configurations and $W$ is the weight function.
From this point, we will omit $\sigma$ and $\beta$ like the last line of eq.(\ref{eq:inner-product}) for simplicity.

The derivative of the inner product $\tilde{A}$ gives us the gauge fixed Berry connection, 
\begin{equation}
\begin{split}
A(\theta) 
\equiv \Braket{\Psi(\theta)|\frac{\partial}{\partial \theta}|\Psi(\theta)}
&= \left.\frac{\partial \tilde{A}(\theta,\phi)}{\partial \phi}\right|_{\theta=\phi} \\
&= \left[N_\theta N_\phi \sum_c \frac{\partial W(c,\theta,\phi)}{\partial \phi}
 + N_\theta \frac{\partial N_\phi}{\partial \phi} \sum_c W(c,\theta,\phi)\right]_{\theta=\phi} \\
&= N_\theta^2 \sum_c \left.\frac{\partial W(c,\theta,\phi)}{\partial \phi}\right|_{\theta=\phi}
 + N_\theta \frac{\partial N_\theta}{\partial \theta} \sum_c W(c,\theta,\theta).
\end{split}
\end{equation}
The complex conjugate of the Berry connection can be obtained in the same way,
\begin{equation}
A^*(\theta) 
\equiv \left(\Braket{\Psi(\theta)|\frac{\partial}{\partial \theta}|\Psi(\theta)}\right)^*
= \left.\frac{\partial \tilde{A}(\theta,\phi)}{\partial \theta}\right|_{\theta=\phi} 
= N_\theta^2 \sum_c \left.\frac{\partial W(c,\theta,\phi)}{\partial \theta}\right|_{\theta=\phi}
 + N_\theta \frac{\partial N_\theta}{\partial \theta} \sum_c W(c,\theta,\theta).
\end{equation}
Since the Berry connection is pure imaginary, the difference between it and its complex conjugate is just twice as much as itself,
so the Berry connection is given by
\begin{equation}
\begin{split}
A(\theta) = \frac{A(\theta)-A^*(\theta)}{2}
&= \left.\frac{1}{2}\left(\frac{\partial}{\partial \theta} - \frac{\partial}{\partial \phi}\right)\tilde{A}(\theta,\phi)\right|_{\theta=\phi} \\
&= \frac{N_\theta^2}{2}
   \sum_c \left.\left(\frac{\partial}{\partial \theta}-\frac{\partial}{\partial \phi}\right)
          W(c,\theta,\phi)\right|_{\theta=\phi}\\
&= \frac{1}{2}\frac{\sum_c \left.\left(\frac{\partial}{\partial \theta}-\frac{\partial}{\partial \phi}\right)W(c,\theta,\phi)\right|_{\theta=\phi}}
        {\sum_c W(c,\theta,\theta)}.
\end{split}
\end{equation}
Unfortunately, this general form cannot be evaluated directly by Monte Carlo method.

Now, we will consider a special case; an antiferromagnetic Heisenberg model with local twist.
In this model, the Hamiltonian of which is
$\mathcal{H(\theta)} = J\sum_{<ij> \ne <kl>}S_i S_j + J S_k^zS_l^z + \frac{J}{2}(e^{i\theta}S_k^+S_l^- + e^{-i\theta}S_k^-S_l^+),$
the weight function can be decomposed to a parameterized phase factor and a weight function without parameters,
$W(c,\theta,\phi) = \exp(i\theta\nuc + i\phi\nlc)W_0(c)$,
where $\nlc$ is the difference between the number of $S_k^+S_l^-$ and that of $S_k^-S_l^+$ at $\tau < \beta/2$
and $\nuc$ is one at $\tau > \beta/2.$
$W_0(c)$ is an abbreviation of $W(c,0,0)$, which is the same as an ordinary Heisenberg model
except for the boundary condition along to the imaginary-time direction
(fixed in the present method and periodic in the ordinary one.)
Finally, we succeed to derive a Monte Carlo evaluable formula for the Berry connection, 
\begin{equation}
\begin{split}
A(\theta) 
&= \frac{\sum_c i\nmc e^{i\theta\npc}W(c,0,0)}{2\sum_c e^{i\theta\npc}W(c,0,0)} \\
&= \frac{i}{2}
\frac{\sum_c \nmc e^{i\theta\npc}W_0(c)}{\sum_cW_0(c)}\biggm/
\frac{\sum_c e^{i\theta\npc}W_0(c)}{\sum_cW_0(c)} \\
&= \frac{i}{2} \frac{\Braket{n^- e^{i\theta n^+}}}{\Braket{e^{i\theta n^+}}},
\end{split}
\end{equation}
where $n^\pm = n^\text{U}\pm n^\text{L}$ and $\braket{O}$ is the expectation value of $O$ over the Monte Carlo simulation with $\theta=0$.

This Monte Carlo expectation form, of course, suffers from a ``complex weight problem.''
This problem arises as the denominator of this will have an exponentially small expectation value
and an constant-order variance
when the projection parameter $\beta$ becomes larger.
Fortunately, at some parameter ($\theta/\pi = p/q$ where $p$ and $q$ are mutually prime and $q$ is even)
the meron cluster algorithm can be applied to this sign problem,
and we can calculate the Berry phase from those discrete data by numerical integration~\cite{MotoyamaT2013}.

\section{Demonstration}
To demonstrate the present method, we calculated the antiferromagnetic Heisenberg model
on a staggered bond-alternating ladder.
The Hamiltonian is
\begin{equation}
  \mathcal{H} = \sum_{j=1}^{L} [ J\left(1 + (-1)^{j}\delta\right) \boldsymbol{S}_{1,j}\cdot\boldsymbol{S}_{1,j+1}  
                                        + J\left(1 - (-1)^{j}\delta\right) \boldsymbol{S}_{2,j}\cdot\boldsymbol{S}_{2,j+1}
                                        +  J' \boldsymbol{S}_{1,j}\cdot\boldsymbol{S}_{2,j} ],
\end{equation}
where $L$ is the ladder length (and so the number of sites is $N=2L$) and $\boldsymbol{S}_{i, j}$ stands for the $S=1/2$ spin operator on the $j$-th site of the $i$-th leg.
The boundary condition along the ladder is periodic, that is, $\boldsymbol{S}_{i,L+1} = \boldsymbol{S}_{i,1}$ $(i=1,2)$.
Depending on the strength of rung coupling the spatial pattern of the valence bond changes.
When rung coupling is weak, $J' < J'_\text{c}(\delta)$, valence bonds are on all strong leg bonds ($J(1+\delta)$ bonds).
Otherwise, $J' > J'_\text{c}(\delta)$, they are on rung bonds.
The threshold $J'_c(\delta)$ is the quantum critical point~\cite{MartinDelgadoSS1996}.
In the present study, we fix $J = 1$ and $\delta = 0.5$,
for which the quantum critical point has been estimated as $J'_\text{c} \sim 1.2$~\cite{Okamoto2003}.

\begin{figure}[t]
\hspace{-0.5em}
\begin{minipage}{0.5\hsize}
  \includegraphics[width=0.9\linewidth]{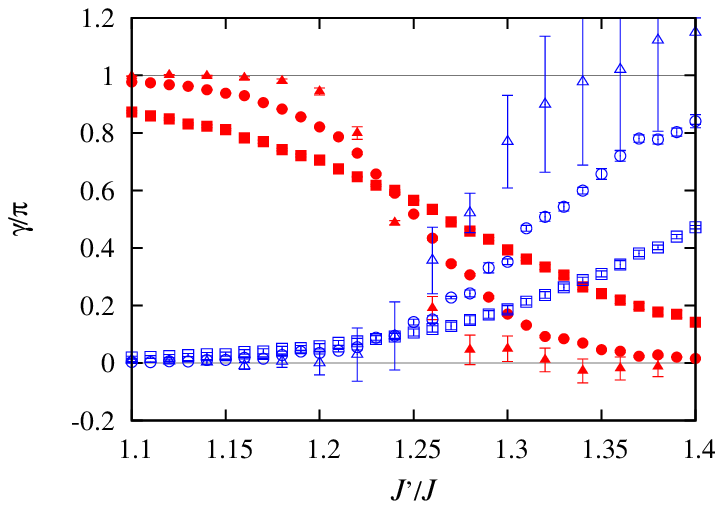}
  \caption{
    The local Berry phase of the staggered ladder
    with system size $L=8$ (squares), $16$ (circles), and $32$ (triangles)
    on the leg bonds (solid red symbols) and the rung bonds (open blue symbols).
    The projection parameter $\beta$ is $2L$~\cite{MotoyamaT2013}.
  }
  \label{fig:bp}
\end{minipage}
\hspace{1em}
\begin{minipage}{0.5\hsize}
\includegraphics[width=.9\linewidth]{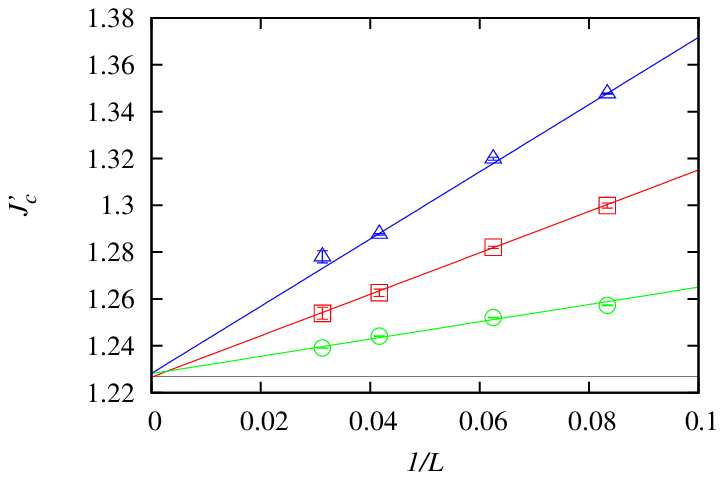}
\caption{
    The estimation of the critical point of staggered ladders obtained by the local $Z_2$ Berry phase
    on the leg bond (circles), rung bond (triangles), and their crossing point (squares).
    The horizontal line, $J'_\text{c}/J = 1.2268,$ is the FSS result of staggered susceptibility~\cite{MotoyamaT2013}.
}
\label{fig:criticalpoint}
\end{minipage}
\end{figure}
Figure~\ref{fig:bp} shows the Berry phase calculation for system sizes $L=8, 16, 32$ and projection parameter $\beta=2L$.
When the rung coupling becomes larger,
the local Berry phase on the strong leg bonds $\gamma^\text{leg}$ and the rung bonds $\gamma^\text{rung}$ varies from $\pi$ to $0$
and from $0$ to $\pi$, respectively.
Since the projection parameter is not so large close to the critical point
that projected states do not reach the ground states, these curves are not step functions.
In this case, however, since the energy gap remains finite except at the critical point, the curves converge to step functions as $L$ and $\beta$ become larger.

To estimate the critical point,
we observed the size dependency of three points:
$J_\text{c}^{\prime\text{leg}}(L)$ and $J_\text{c}^{\prime\text{rung}}(L)$ are the points where $\gamma^\text{leg} = \pi/2$ and $\gamma^\text{rung} = \pi/2 $, respectively,
and $J_\text{c}^{\prime\text{cross}}(L)$ is the one where  $\gamma^\text{leg} = \gamma^\text{rung}$.
The critical point in the thermodynamics limit, $J'_\text{c},$ is estimated
by size extrapolation of $J'_\text{c}(L)$ for lattice sizes up to $L=32$;
$J_\text{c}^{\prime\text{leg}} = 1.2281(18),$ $J_\text{c}^{\prime\text{rung}} = 1.2282(18)$, and $J_\text{c}^{\prime\text{cross}} = 1.2266(6)$ (Fig.~\ref{fig:criticalpoint}).
These results are consistent within statistical errors with the independent finite-size scaling (FSS) analysis for the staggered susceptibility, $J'_\text{c}/J = 1.2268(2)$.

\begin{figure}[t]
\hspace{-0.5em}
\begin{minipage}{.5\hsize}
\includegraphics[width=.9\linewidth]{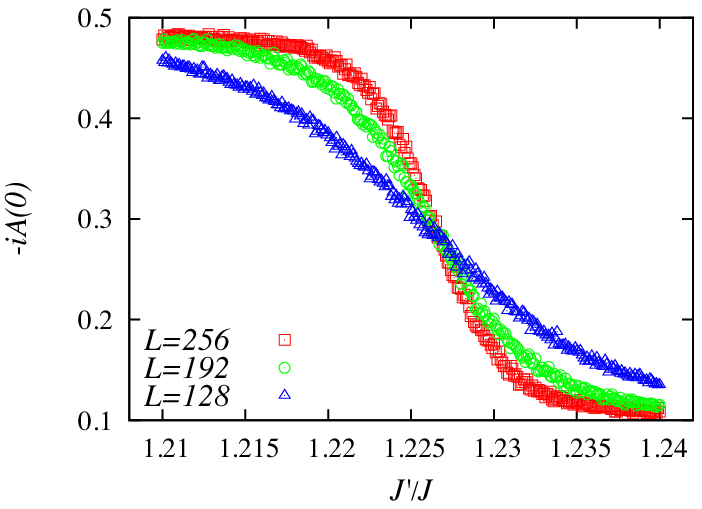}
\caption{
  The leg twist gauge-fixed Berry connection 
  of the staggered ladder with system sizes $L = 128, 192, 256$ and projection parameter $\beta = 2L.$
  The crossing point is $J' = 1.227(1). $
}
\label{fig:bc}
\end{minipage}
\hspace{1em}
\begin{minipage}{.5\hsize}
\includegraphics[width=.9\linewidth]{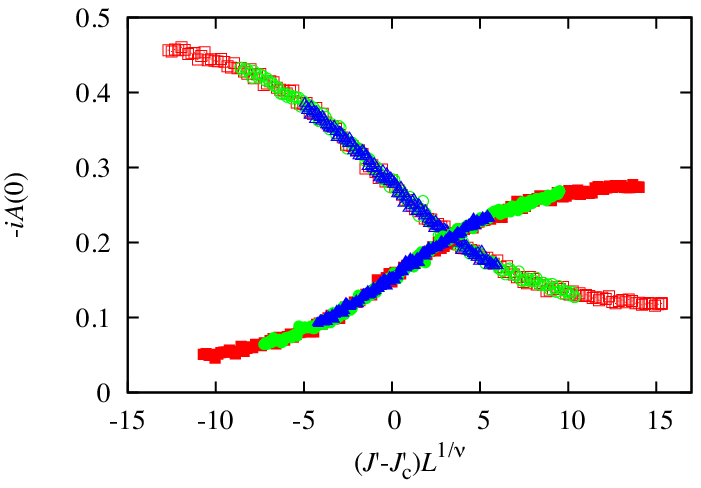}
\caption{
  The finite size scaling plot for the imaginary part of the gauge-fixed Berry connection
  of the staggered ladder with system sizes $L = 128, 192, 256.$
  The downward to the right curve and the upward one are the leg twist and the rung twist Berry connection, respectively.
  The fitting parameters are $(J'_c, \nu) = (1.2268(1), 0.738(8))$ for the leg twist and $(1.2265(1), 0.747(10))$ for the rung twist.
  ~\cite{MotoyamaT2013}.
}
\label{fig:bc-fss}
\end{minipage}
\end{figure}
Figure~\ref{fig:bc} shows the leg twist gauge-fixed Berry connection with twist angle $\theta = 0.$
It is clear that the three curves, $L=128,192,256$, cross at one point, $J' = 1.227(1).$
Under the gauge transformation, $|\psi(\theta)\rangle \to e^{i\chi(\theta)} |\psi(\theta)\rangle,$
the Berry connection varies as 
$A(\theta) \to A(\theta) + i\partial_\theta\chi(\theta)$, where $\chi$ is some arbitrary periodic real function such as $\chi(\theta) = \chi(\theta+2\pi)$.
This means that the gauge transformation shifts the $J'$-$A$ curve only vertically by a constant
and does not change the crossing point.
Thus, the permitted finite size scaling is only $J'$ rescaling, such as
$A(J',N) = f((J'-J'_\text{c})N^{1/\nu})$, with some universal function $f$.
Figure~\ref{fig:bc-fss} shows the result of the finite size scaling result
of the gauge-fixed Berry connection at $\theta=0$.
The system sizes are $L=128, 192, 256$ and the projection parameter is $\beta = 2L.$
The scaling parameters are 
$J'_\text{c} = 1.2268(1)$ and $\nu = 0.738(8)$ for the leg twist
and $J'_\text{c} = 1.2265(1)$ and $\nu = 0.747(10)$ for the rung twist.
These estimates agree with the result of the finite size analysis of staggered susceptibility
obtained by loop algorithm.

\section{Conclusion}
We presented the Monte Carlo form of the gauge-fixed Berry connection
and used it to calculate the local $Z_2$ Berry phase.
We also proposed that the gauge-fixed Berry connection can be used as
an effective tool to catch the quantum phase transition.
For the demonstration, we applied these to the antiferromagnetic Heisenberg model
on a staggered bond-alternating ladder.
The estimated critical point is consistent with the finite size analysis of staggered susceptibility.

We used ALPS libraries~\cite{ALPS2011s,ALPSweb} to develop the simulation code
and ALPS application (ALPS/looper~\cite{TodoK2001,LOOPERweb}) to calculate the staggered susceptibility to check our result.
We also used BSA~\cite{Harada2011, BSAweb} for Bayesian finite size analysis.
We acknowledge support by
KAKENHI (No.~23540438),
JSPS,
Grand Challenges in Next-Generation Integrated Nanoscience,
Next-Generation Supercomputer Project,
the HPCI Strategic Programs for Innovative Research (SPIRE),
the Global COE program ``the Physical Sciences Frontier,''
MEXT, Japan,
and the Computational Materials Science Initiative (CMSI).


\begin{thebibliography}{10}

\bibitem{Hatsugai2004}
Y.~Hatsugai: J. Phys. Soc. Jpn. {\bfseries 73} (2004) 2604.

\bibitem{MotoyamaT2013}
Y.~Motoyama and S.~Todo: Phys. Rev. E {\bfseries 87} (2013) 021301(R).

\bibitem{MartinDelgadoSS1996}
M.~A. Mart\'in-Delgado, R.~Shankar, and G.~Sierra: Phys. Rev. Lett. {\bfseries
  77} (1996) 3443.

\bibitem{Okamoto2003}
K.~Okamoto: Phys. Rev. B {\bfseries 67} (2003) 212408.

\bibitem{ALPS2011s}
B.~Bauer and {\it et al.}: J. Stat Mech.  (2011) P05001.

\bibitem{ALPSweb}
{http://alps.comp-phys.org/}.

\bibitem{TodoK2001}
S.~Todo and K.~Kato: Phys. Rev. Lett. {\bfseries 87} (2001) 047203.

\bibitem{LOOPERweb}
{http://wistaria.comp-phys.org/alps-looper/}.

\bibitem{Harada2011}
K.~Harada: Phys. Rev. E {\bfseries 84} (2011) 056704.

\bibitem{BSAweb}
{http://kenjiharada.github.io/BSA/}.

\end{thebibliography}
\end{document}